# Performance Analysis of View Maintenance Techniques for DW

[1]Prakasha S. and [2]R. Selvarani

*Abstract*

**A Data Warehouse stores integrated information as materialized views over data from one or more remote sources. These materialized views must be maintained in response to actual relation updates in the remote sources. The data warehouse view maintenance techniques are classified into four major categories self maintainable recomputation, not self maintainable recomputation, self maintainable incremental maintenance, and not self maintainable incremental maintenance. This paper provides a comprehensive comparison of the techniques in these four categories in terms of the data warehouse space usage and number of rows accessed in order to propagate an update from a remote data source to a target materialized view in the data warehouse.**

## I. INTRODUCTION

A DWH stores integrated information over data from one or more remote data sources for query and analysis [11]. The integrated information at the data warehouse is stored as materialized views. A materialized view is the result relation of the evaluation of the relational algebra expression that defines the view relation [3]. Using these materialized views, user queries can be answered quickly as the information may be directly available or can be calculated.

A problem known as the view maintenance problem is how to maintain the materialized views so that they can be kept up to date in response to updates of the actual relations in the remote data sources. . The database systems understand view management and view definitions and know what data is needed for propagating updates to the views.

In a data warehouse, the query expressions that define views and actual relations may be stored at many sites. The sources may inform the data warehouse when an update occurs but they might not be able to determine what data is needed for updating the views at the data warehouse. Therefore they may send only the actual data updates or the entire updated relations to the data warehouse [16]. Upon receiving this information, the data warehouse may find that it needs some additional source data in order to update the views. Then it will issue some queries to some of the sources to request the additional source data. Some of the sources may have updated their data again before they evaluate the requesting queries from the data warehouse. Therefore they will send incorrect additional data to the data warehouse, which subsequently will use the incorrect data to compute the views. This phenomenon is called distributed view maintenance anomaly[17]. Solving the view maintenance problem in data warehouses is thus more complicated than that in traditional database systems. The objectives of this paper are to provide a classification of different view maintenance techniques that have been proposed and to conduct a comprehensive comparison of these techniques in terms of space usage and number of rows accessed using the TPC benchmark for decision support queries.

## II. CLASSIFICATION OF DWH VM TECHNIQUES

The existing DWH VM techniques can be classified into two broad categories: RVM and IVM. Depending on whether the data warehouse has to query the remote data sources in order to calculate the new views, the techniques can be further classified as self-maintainable or not self-maintainable. The below subsections discuss these four categories.

### A. The Self-Maintainable Recomputation Category

Materialized views can be computed by using the view definitions and other materialized views at the data warehouse. The current materialized views being maintained have no contribution to the calculation of the new views. Some techniques replicate all or part of the remote data at the data warehouse. We can view these replicated data as some kind of materialized views at the data warehouse. Others such as the self-maintenance warehouse approach discussed in [18] store the remote relations at the data warehouse as additional materialized views to provide data needed when the data warehouse computes the new views. Therefore, the data warehouse will never have to query the data sources for additional data.

A self-maintainable materialized V view can be defined in two ways. In first case, the view V is defined as

$$V = \Pi_{proj}(\sigma_{cond}(v_1 \bowtie v_2 \bowtie \ldots \bowtie v_N))$$

where all vi 's are self-maintainable materialized views stored in the data warehouse[10]

However, a self-maintainable data warehouse view cannot be defined as

$$V = \Pi_{proj}(\sigma_{cond}(v_1 \bowtie v_2 \bowtie \ldots \bowtie v_k \bowtie r_{k+1} \bowtie r_{k+2} \bowtie \ldots \bowtie r_N))$$

"This work was supported by DSI under encouraging Research Activities Performance Analysis of View Maintenance Techniques for DW

Prakasha S is with Dept of Computer Science, Dayananda Sagar College of Engineering since 2007 july Shavige Malleshwara Hills, Kumara Swamy Layout Bangalore , Karnataka, India
e mail: sprakashjpg@yahoo.co.in
R.Selvarani is with Dept of Computer Science, Dayananda Sagar College of Engineering since 2007 july Shavige Malleshwara Hills, Kumara Swamy Layout Bangalore , Karnataka, India
E Mail:- selvss@yahoo.co.in.

where all relations vi 's are self-maintainable materialized views residing at the data warehouse, all relations ri 's are self-maintainable relations residing at the remote data sources, and there are totally N relations in the definition of view V. The reason is as follows. The data warehouse is still self-maintainable when an update of relation rl is propagated to the data warehouse as all vi 's are available in the data warehouse. However, when an update of view vi is propagated to view V, the relation rl is not available at the data warehouse. Therefore, the data warehouse has to send a query to the remote data sources to get the relation rl in order to calculate view V. Thus view V is not self-maintainable[12].

An advantage of the techniques in this category is that the view maintenance anomaly problem is avoided as all necessary data are available at the data warehouse. The data warehouse knows the view definitions and what to do with the views to keep them up to date. It eliminates accesses to the remote relations, and therefore, it does not compete with the remote data sources' local resources. Extra storage and time are thus needed to maintain these additional views.

B. The Not Self-Maintainable Recompilation Category

When an update occurs at the data source or periodically, the source will inform the data warehouse. According to the query expression that defines the view, the data warehouse may get part of data it wants from other materialized views at the data warehouse, and issue queries to the sources to get the other data it does not have. The sources send the query results back to the data warehouse. Based on the query results, the data warehouse calculates the views and stores the results as materialized views in the data warehouse.

The DWH may replicate part of the remote relations in the warehouse. However, these data are not enough for maintaining the materialized views. Therefore, the data warehouse will have to query the remote data sources for additional data in order to maintain the views. An extreme case is where the data warehouse does not replicate any remote relations.

If the view maintenance process is not designed carefully, the distributed view maintenance anomaly problem will occur. Suppose that there is a data warehouse system where the remote data sources send updated relations to the data warehouse whenever an update occurs at the data sources [14]. Upon receiving the information, the data warehouse is ready to compute the new views. But now let us assume that the data warehouse finds that it needs some other relations at some remote data sources to compute the new views. It will issue queries to these data sources [15]. Suppose the data sources that sent the updated relations to the data warehouse update the relations again before they receive the queries from the data warehouse. The data sources answer the query and send the results to the data warehouse. These results might contain extra information that is incorrect. The data warehouse will then use the incorrect data to compute the new views, which will result in incorrect new views.

C. The Self-Maintainable Incremental Maintenance Category

In this category, the DWH views are maintained by using the view definitions, the materialized views, and the view updates. The DWH will never query the remote data sources as the information at the data warehouse is enough for maintaining the views. The data warehouse computes the view updates, then adds them to the materialized views. The process is incremental. Normally, only necessary remote relations, or views of the remote relations are stored at the data warehouse as materialized views. In the extreme case, all remote relations can be replicated at the data warehouse. The self-maintainable warehouse approaches discussed in [4] , [2] and [10] belong to this category.

Let us discuss how to maintain a view V that is defined as $V = \Pi_{proj}(\sigma_{cond}(v_1 \bowtie v_2 \bowtie \ldots \bowtie v_N))$ where each vi is a materialized view and is defined as either

$$v_i = \Pi_{proj}(\sigma_{cond}(v_{11} \bowtie v_{12} \bowtie \ldots \bowtie v_{1N}))$$

where each v1j is a view defined by other auxiliary materialized views,

Finally, at the lowest level of the view hierarchy discussed earlier in this paper, view vMj can only be defined by relations at the remote data sources as follows:

$$v_{Mj} = \Pi_{proj}(\sigma_{cond}(r_{M1} \bowtie r_{M2} \bowtie \ldots \bowtie r_{MN})) \text{ where each } r_{Mj} \text{ is a base relation.}$$

The above view V is thus defined by M levels of the materialized views in the view hierarchy. In the second case, the view can only be defined by base relations r1j.

All intermediate materialized views can be viewed as auxiliary views. These auxiliary views are self-maintainable. The materialized view V is self-maintainable by using the update information and additional information from the auxiliary views. The data warehouse views, including views such as V and auxiliary views, can be maintained starting with those views that do not depend on any other auxiliary views, working up to the final original view V.

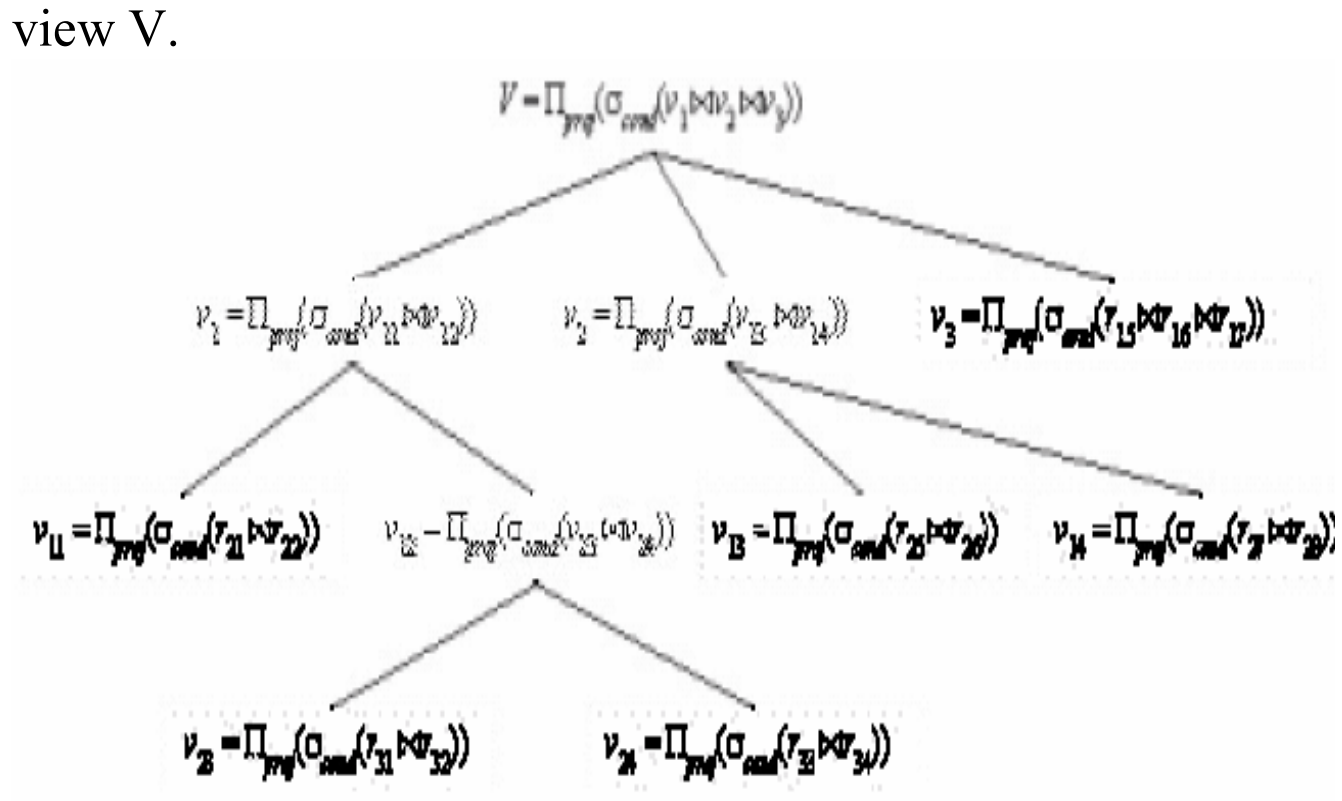

figure 1: View Hierarchy Example.

All related materialized primary views, auxiliary views and base relations can be drawn in a hierarchy structure as shown in Figure 1. All leaves in the hierarchy structure are those materialized views defined by the base relations. In this example, V is the primary materialized view. Views v1, v2 and v12 are materialized auxiliary views defined by other materialized auxiliary views. Views v3, v11, v13, v14, v23 and v24 are materialized auxiliary views defined by the base relations. All relations rij 's are the base relations. The views in the leaves should be maintained first. Suppose an update for r33 occurs in the data source. View v24 should be maintained first. Then views v12 and view v1 must be maintained next in that order. Finally, the primary view V is maintained.

The data warehouse never needs to query the remote data source to get additional data. The data warehouse maintenance operations can be totally separated from other OLTP operations[8]. Whether the remote data source is available or not will not affect the data warehouse view maintenance process. However, in order to make the materialized views self-maintainable, the auxiliary views are stored in the data warehouse to provide the additional information. Extra storage and time overhead are therefore required to maintain the auxiliary views themselves. [10].

A. The Not Self-Maintainable Incremental Maintenance Category

The DWH has to query the remote data sources whenever necessary because the information at the data warehouse is not enough to maintain the view. A number of existing approaches fall under this category. Among them are the unrestricted base access [14] and runtime warehouse self-maintenance [5]

a) Unrestricted Base Access

In the Unrestricted Base Access approach [14], the data warehouse accesses the actual relations from the data sources whenever necessary in order to maintain the materialized views. There are many proposed algorithms that follow this approach. The Eager Compensating Algorithm (ECA) is the simplest among them. It is also the fastest algorithm that will let the data warehouse remain in a consistent state [14]

The data warehouse keeps a temporary table called COLLECT to keep the intermediate answers it receives from the data sources. It also keeps a set called Unanswered Query Set It then creates a temporary COLLECT table and UQS set for processing this specific query, and sets both the COLLECT table and UQS to empty.

The data warehouse writes the query Qi to the UQS and sends the query Qi to the data source. Suppose there is another update Uj that occurs at the same data source. The data source sends the update Uj to the data warehouse before it receives the query Qi. The data warehouse now receives the update Uj. It knows that the upcoming answer for Qi from the data source will contain extra information caused by simultaneous Uj update at the data source[15].

This approach calculates view updates then adds them to the old views in order to get the new views. However, the data warehouse has to access N –1 remote source actual relations in order to propagate one source update.

In this approach, the data warehouse may have to send queries back to the sources and waits for answers in order to compute the view updates. Therefore, this approach has the same limitation as the not self-maintainable recomputation approach. Computing these queries consumes remote sources' local resources, and will slow down other OLTP operations. If the remote sources are unavailable, the data warehouse will not get the answers it needs.

c) Runtime Warehouse Self-Maintenance

Design-time self-maintainability is not flexible. It may be difficult or impossible for us to know the exact contents of the views and their updates at design time. To solve this problem, a run time warehouse maintenance approach has been introduced [5].

The basic idea of the runtime self-maintenance approach is that the data warehouse generates the self-maintainable test for the views to determine whether the views are self-maintainable for a particular update. At run time, the self-maintainable test determines the views for self-maintainability [1].

. If the view is not self-maintainable, then the data warehouse has to query the remote data sources for those relations it needs in order to update the view. In this case, this approach is similar to the unrestricted base access approach.

## III. PERFORMANCE ANALYSIS

We conduct an analysis to compare the performance of different algorithms in the four categories. We consider only the problem of single view maintenance in a single source environment because the ECA algorithm in the not self-maintainable incremental maintenance category can only be used in this environment [14].

A. Performance Measurements

In our analysis, only Select-Project-Join views are considered. We measure the performances of the techniques in terms of space and number of row accesses, which are defined as follows:

- Space: total space needed to store the data in the data warehouse, including space for auxiliary views. We do not consider indices.
- Number of rows accessed: the number of rows that must be accessed in the data warehouse and the data sources in order to integrate the updates into the data warehouse.

B. Comparison Based on Space Needed in the Data Warehouse

a) Self-Maintainable Recomputation

The techniques in this category do not query the remote data source for additional data in order to maintain the data warehouse materialized views. The data warehouse can replicate all or part of the remote base relations at the data warehouse. These additional data take space at the data warehouse. Here we consider the case where the materialized views are defined by other materialized views (auxiliary views) at the data warehouse, and all auxiliary views are replicated remote relations. A view V is defined as

b) Not Self-Maintainable Recomputation

Here we consider only the case where the data warehouse does not replicate any base relations. Therefore, the data warehouse always has to query the remote data sources. The data warehouse stores only materialized views. In this extreme situation, the amounts of space needed in the best case, the average case and the worst case are the same, and are equal to Card(V) ts(V).

b) Self-Maintainable Incremental Maintenance

Similar to the self-maintainable recomputation techniques, the techniques in this category can replicate all or part of the remote data at the data warehouse. Here we consider only the case where the materialized views are defined by other materialized views (auxiliary views) at the data warehouse, and all auxiliary views are replicated remote relations [6].

c) Not Self-Maintainable Incremental Maintenance

Here we consider the Eager Compensating Algorithm (ECA) for this category. In ECA, a temporary table COLLECT is used to store intermediate query answers. For every update, the queries including compensated queries are sent to the data source. Note that the COLLECT table is empty only when there is no query to the data source, or the answers for all the queries are returned to the data warehouse before a new update occurs at the data source. This is the best case[9].

C. Comparison Based on the Number of Rows Accessed

To analyze the number of rows accessed at the data warehouse by the techniques, we made the following assumptions:

- The set of a primary view and its auxiliary views (if any) is independent to other sets of primary views and their auxiliary views.
- We do not consider indices. Linear search is thus used to check if a record satisfies a select or join condition.
- All auxiliary views are self-maintainable and are replicated base relations.
- Updates to auxiliary views and primary views are for appending only.

a) Self-Maintainable Recomputation

The data warehouse will never query the remote data sources as all necessary data are available at the data warehouse. Updates from the remote data sources have to be propagated to the replicated relations at the data warehouse first, then the data warehouse recalculates the view relation and stores the result at the data warehouse as the new materialized view. In order to propagate an update to data warehouse replicated relation, the number of rows to be accessed at the data warehouse is the cardinality of the relation itself plus the cardinality of the update. That is, Card(r) + Card(U).

b) Not Self-Maintainable Recomputation

Only source data is required to be accessed. The reason is that the warehouse recalculates the full view using the source data each time. It does not use the data warehouse data. Suppose the system locks all base relations in order to evaluate the query expression that defines the view. If the nested-loop join method [3] is used to evaluate it, the total number of rows to be accessed is Card(r)N.

Another strategy such as the one described in can also be used to evaluate the query Q that defines the view V[14]. It will reduce the total number of rows to be accessed. Let us rename the actual relations according to the join order.

c) Self-Maintainable Incremental Maintenance

No queries are sent to the data sources for additional information. Therefore, the number of rows accessed in the data source is equal to 0. For N base relations in a view, Nav should be less than or equal to N. In the worst case, Nav is equal to N.

At first, the auxiliary view itself has to be maintained before the primary materialized view can be maintained. Let Card(U) stand for the cardinality of update U. According to our assumption that auxiliary views are self-maintainable and updates are used for appending only, the number of rows needed to be accessed in order to maintain the auxiliary view is Card(U) + Card(AV). Let Card( AV) stand for the cardinality of the auxiliary view update, which is the same as Card(U). Then the update is propagated to the primary view[8]. We need to calculate the primary view update.

d) Not Self-Maintainable Incremental Maintenance

In the ECA algorithm, all tuples in the view table have to be accessed in order to find a tuple to integrate with the view update. However, the data warehouse may have to access data from remote sites except for the best case[11]. Parts of these queries are compensated. we derive the number of wrapper queries corresponding to queries with N – n relations in the multiple tuple update case .

## IV. CONCLUSIONS AND FUTURE WORK

All data warehouse view maintenance techniques can be classified into four major categories. They are self-maintainable recomputation, not self-maintainable recomputation, self-maintainable incremental maintenance, and not self-maintainable incremental maintenance. Their advantages and disadvantages are l

Both self-maintainable recomputation and self maintainable incremental maintenance approaches totally separate the data warehouse view maintenance operations from the OLTP operations. Therefore, the view maintenance operations will not consume data sources' local resources. These operations only consume the data warehouse's resources. Even if the remote data sources are not available, the data warehouse view maintenance process can continue running. However, a part or all source data are replicated at the data warehouse to make the data warehouse view maintenance process self-maintainable. These replicated data take space. Data transfer processes are implemented to transfer data from the remote data sources to the data warehouse. Design, implement and maintain these processes are time-consuming. A lot of unnecessary data may be duplicated at the data warehouse. However, these are the approaches that probably many large companies have to take if they want to separate their data warehouse view maintenance operations from their OLTP operations.

TABLE I
ADVANTAGES AND DISADVANTAGES OF THE VM TECHNIQUES

| Category | Advantage | Disadvantage |
|---|---|---|
| Self-Maintainable Recomputation | -Data warehouse view maintenance operations are totally separated from OLTP operations;<br>-Unavailable source will not block the data warehouse view maintenance process; | -Data are replicated at data warehouse;<br>-Need extra data storage for replicate data;<br>-Have to implement and maintain data transfer processes to transfer data from sources to data warehouse; |
| Not Self-Maintainable Recomputation | -Very simple to implement;<br>-No replicate data at the data warehouse;<br>-No extra data storage for replicate data;<br>-Do not have to implement and maintain data transfer processes to transfer data from sources to data warehouse; | -Unavailable source will block the data warehouse view maintenance process;<br>-Evaluating queries at the data sources consumes local resources;<br>-Data warehouse view maintenance operations are not separated from OLTP operations; |
| Self-Maintainable Incremental Maintenance | -Data warehouse view maintenance operations are totally separated from OLTP operations;<br>-Unavailable source will not block the data warehouse view maintenance process;<br>-In the worst case, the number of rows accessed to maintain a view is the lowest; | -Data are replicated at data warehouse;<br>-Need extra data storage for replicate data;<br>-Have to implement and maintain data transfer processes to transfer data from sources to data warehouse; |
| Not Self-Maintainable Incremental Maintenance | -No replicate data at the data warehouse;<br>-No extra data storage for replicate data;<br>-Do not have to implement and maintain data transfer processes to transfer data from sources to data warehouse; | -Unavailable source will block the data warehouse view maintenance process;<br>-Evaluating queries at the data sources consume local resources;<br>-Data warehouse view maintenance operations are not separated from OLTP operations;<br>-Have to design the view maintenance process carefully to avoid the anomaly problem;<br>-In the worst case the number of rows accessed is the highest;<br>-Performance is down-graded rapidly;<br>-Need extra storage for intermediate data (COLLECT tables); |

Both the not self-maintainable recomputation and not self-maintainable incremental maintenance approaches suffer from some common disadvantages. As the remote data sources have to process queries from the data warehouse that consume their limited local resources, the OLTP system will be slow. Once a data source is unavailable, the data source will not be able to answer queries sent from the data warehouse in time. It will block the data warehouse view maintenance process. The not self-maintainable incremental maintenance approach has some additional disadvantages. To avoid the anomaly problem, the view maintenance process must be designed carefully. If a lot of updates happen at the data sources, the data warehouse may issue many compensating queries. It is very possible that the data warehouse may never get the final query results. Both approaches also have some common advantages. As there is no replicate data stored at the data warehouse, no data transfer process has to be implemented and maintained.

There is no extra space for storing replicate data. Both approaches are good for small to mid-sized companies whose OLTP database systems are not too busy. Among all the four categories, self-maintainable incremental maintenance is the best in terms of space used in the data warehouse and number of rows accessed in order to propagate an update to the target materialized view in the data warehouse[13] . As the cost of data storage becomes increasingly low, this is the best approach to implement a data warehouse.

## VI. Biographies

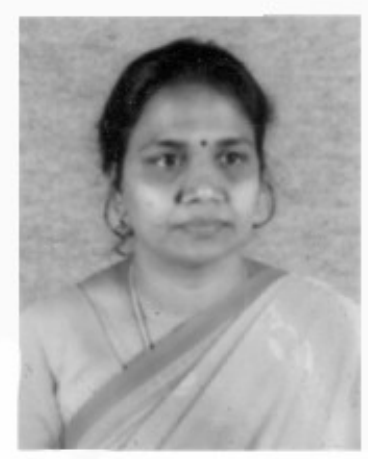

R.Selva Rani born in Coimbatore she did masters from manipal university and pursuing Ph.D and JNT Hyderabad. Her area of interest are Data Warehousing & Data Mining, DBMS, Software Engineering.

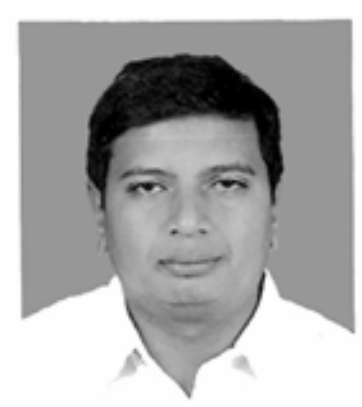

Prakasha S born in Bangalore 01-1-1972, he did masters B.E from SJCE Mysore university, and M.Tech from MSRIT, VTU Belgaum, his area of interest are Data Warehousing & Data Mining, DBMS, Software Engineering.